\documentclass[runningheads]{llncs}
\usepackage{booktabs}
\usepackage{graphicx}
\usepackage{xcolor}
\usepackage{algpseudocode}
\usepackage{algorithm}

\usepackage{academicons}
\usepackage{xcolor}
\begin{document}
\setcounter{secnumdepth}{5}
\title {LSTM-based QoE Evaluation for Web Microservices’ Reputation Scoring}
%
%

\author{
Maha Driss\inst{1,2}\orcidID{0000-0001-8236-8746}} 

\authorrunning{M. Driss.}

\institute{Computer Science Department, CCIS, Prince Sultan University, Saudi Arabia \and
RIADI Laboratory, ENSI, University of Manouba, Tunisia}

\maketitle            
\begin{abstract}
Sentiment analysis is the task of mining the authors' opinions about specific entities. It allows organizations to monitor different services in real time and act accordingly. Reputation is what is generally said or believed about people or things. Informally, reputation combines the measure of reliability derived from feedback, reviews, and ratings gathered from users, which reflect their quality of experience (QoE) and can either increase or harm the reputation of the provided services. In this study, we propose to perform sentiment analysis on web microservices reviews to exploit the provided information to assess and score the microservices' reputation. Our proposed approach uses the Long Short-Term Memory (LSTM) model to perform sentiment analysis and the Net Brand Reputation (NBR) algorithm to assess reputation scores for microservices. This approach is tested on a set of more than 10,000 reviews related to 15 Amazon Web microservices, and the experimental results have shown that our approach is more accurate than existing approaches, with an accuracy and precision of 93\% obtained after applying an oversampling strategy and a resulting reputation score of the considered microservices community of 89\%.


\end{abstract}
\section{Introduction}
In the current era, many customer reviews are available on different platforms and applications: e-commerce, Web services, games, social networks, etc. What interests us in this paper are the web microservices-based applications. A web microservice is a tiny, self-contained component of an online application that performs a specific function or task. The microservices architecture is a methodology for developing software systems consisting of loosely coupled, independently deployable services \cite{wolff2016microservices}. Customers post reviews online as feedback on microservices they have purchased, used, or experienced. These reviews are one of the most effective ways to motivate and encourage potential customers to use services. They reflect users' quality of experience (QoE), which can influence potential customers' perceptions. Positive reviews can enhance the microservice's reputation and encourage new users to try it out, while negative reviews can harm its reputation and discourage potential users. The main issue with these reviews is that they may be ambiguous and unclear, and this is due to various factors such as attitude, emotions, used vocabulary, and previous experiences of the customer. To solve this issue, sentiment analysis techniques \cite{BIRJALI2021Comprehensive} are employed to automatically transform these unstructured reviews into structured data that can be extremely valuable for commercial concerns like reputation management. Having positive reviews and a good reputation as a service can play an important role in its success. It helps attract customers' attention and interest and establish trust and confidence in the service. In this paper, we aim to perform sentiment analysis techniques on web microservices’ reviews to exploit the provided information for services’ reputation assessment and scoring. Our proposed approach is designed and implemented to mine microservices' reviews by categorizing them into different polarity labels and providing a score that is used to measure the microservices' community reputation. This approach applies a deep learning-based sentiment classification that performs the Long Short-Term Memory (LSTM) model \cite{Yu2019review} and employs the Net Brand Reputation (NBR) algorithm \cite{Bilecki2017trust} to assess reputation scores for concerning microservices. This work makes a significant contribution by leveraging the outputs of the LSTM model to classify reviews as positive or negative. These results are then utilized to calculate the overall reputation score of the microservices' community provider through the application of the NBR algorithm. The proposed approach is tested on a set of 10,000 reviews related to 15 Amazon Web microservices. The experimental results have shown that our approach is more accurate than existing approaches, with an accuracy and a precision of 93\% after applying oversampling strategy and a resulting reputation score of 89\%.
The remainder of this paper is structured as follows: Section 2 provides a brief background about Web microservices, sentiment analysis, and reputation assessment. Section 3 presents pertinent related works that implement sentiment classification and reputation assessment for Web microservices. Section 4 details the proposed approach. Section 5 illustrates the implementation of the proposed approach and discusses the experiments that are conducted to test and validate this approach. Section 6 presents the concluding remarks and future works.
\section{Background}
This section presents fundamental concepts related to Web microservices, sentiment analysis, and reputation management.
\subsection{Web Microservices}
A web microservice is a tiny, self-contained component of an online application that performs a specific function or task. The microservices architecture is a methodology for developing software systems consisting of loosely coupled, independently deployable services \cite{surianarayanan2019essentials,driss2022ws}. Each microservice is often responsible for a specific business function and connects with other services through common web protocols. Online microservices are frequently employed to develop sophisticated web systems that demand scalability, fault tolerance, and flexibility. By splitting a web application into smaller, more manageable services, developers may work on each component individually, making it easier to update, test, and deploy changes. The quality of service characteristics (e.g., response time, availability, scalability, security, usability, etc.), which are provided by these Web microservices, have become a primary concern for the users as well as the providers \cite{driss2022req}. One way to improve these characteristics is to analyze the feedback generated by users’ reviews. Mining users’ feedback is crucial since it reflects the service's reputation and leads to its improvement. It generally gives an idea of whether users like the microservice, and if the users do not like it, it indicates what factors contributed to this negative feedback.
\subsection{Web Microservices and Reputation Management}
According to the Concise Oxford Dictionary \cite{baldick1996concise}, "Reputation is generally said or believed about a person’s or thing’s character or standing”. Informally, reputation combines the measure of reliability derived from feedback, reviews, and ratings gathered from users in a certain society. 
The QoE and the reputation of web microservices are closely related. A positive quality of experience can lead to a strong reputation, while a negative quality of experience can harm the reputation of the microservice. When users have a positive experience while using a web microservice, they are more likely to recommend it to others and leave positive reviews or feedback. This can help to build the microservice's reputation and attract new users. On the other hand, if users have a negative experience while using a web microservice, they may leave negative reviews or feedback, which can harm the microservice's reputation.
A reputation model \cite{wahab2015survey} in the context of Web microservices is a method that enables decision-makers to distinguish good and satisfying services from bad and poor ones based on users’ feedback and reviews. In this context, the importance of reputation is derived from the need to help users and service providers to distinguish the quality of the functionalities and performances among similar services based on these services’ history of use and how they behaved in the past. 
\subsection{Sentiment Analysis}
Sentiment Analysis (SA) is defined as analyzing authors' opinions, emotions, or attitudes about specific entities such as products, services, events, and individuals \cite{saberi2017sentiment}. These entities are most likely to be covered by users’ reviews. Sentiment analysis is a process that aims to classify sentiments, and that consists of three different steps \cite{saberi2017sentiment}: 1) sentiment identification, 2) feature selection, and 3) sentiment classification. The input of this process is a dataset of users’ reviews; the output is a set of sentiment polarities (i.e., positive/negative/neutral or positive/negative). There are three main classification levels for SA \cite{yue2019survey}: document-level, sentence-level, and aspect-level SA. In this paper, we tackle the second class of SA since considered users’ opinions will be grouped into a single document that will be analyzed at the sentence level to determine users’ orientations.

\section{Related Works}
Many statistical, fuzzy-logic, and data mining-based approaches for computing web service reputation have been proposed in the literature. These are the most recent and relevant related works.

In \cite{papadakis2019collaborative}, the authors presented a collaborative Service Level Agreement (SLA) and Reputation-based Trust Management (RTM) solution for federated cloud environments. The SLA service explicitly set performance standards and evaluated the real performance of cloud applications installed. Based on the SLA, the collaborative solution's RTM service utilized many technical and user experience parameters to calculate the cloud providers' dependability and customers' trust. The collaborative approach was demonstrated and proven in a genuine federated setting. 
The study, presented in \cite{hasnain2020evaluating}, uses a trust prediction and confusion matrix to rank web services based on throughput and response time. For a benchmark web services dataset, AdaBoostM1 and J48 classifiers were utilized as binary classifiers. The confusion matrix was used to compute trust scores. Correct prediction of trustworthy and untrusted web services has enhanced the overall selection process in a pool of comparable web services. Kappa statistics values were used to evaluate the suggested method and compare the performance of AdaBoostM1 and J48 classifiers. \cite{hasnain2022machine} discussed web service selection utilizing a well-known machine learning technique, REPTree, to forecast trustworthy and untrusted services correctly. Using web services datasets, the performance of REPTree is compared to that of five machine learning models. The authors tested web services datasets using a ten k-fold cross-validation approach. They utilized performance measures, like sensitivity and specificity measures, to assess the effectiveness of the REPTree classifier. The evaluation results of the suggested web services selection technique showed a link between the final selection and the recommended web service trust score. The authors in \cite{al2022cbilstm} presented a reputation-based trust assessment technique using online user evaluations to combine the NBR measure with a deep learning-based sentiment analysis model called CBiLSTM. The suggested deep learning model combined the layers of Convolutional Neural Networks (CNN) and Bidirectional Long Short-Term Memory (BiLSTM). The CNN layers coped with the high dimensionality of text inputs, and the BiLSTM layer investigated the context of the derived features in both forward and backward directions.

The existing works using a reputation-based selection of web services have several limitations, including:
\begin{itemize}
    \item Limited scope: Reputation-based selection approaches typically rely on feedback from a small subset of users, which may not be representative of the broader user community. This can result in biased or incomplete reputation scores.
    \item Difficulty in interpretability: Deep learning-based solutions are often complex and difficult to interpret, making it difficult to understand how they arrive at their reputation scores. This can limit the transparency of the reputation assessment process.
    \item Computational requirements: Hybrid deep learning models used for reputation assessment can be computationally intensive and require significant resources to train and evaluate. This can make them less suitable for use in resource-constrained environments, such as on mobile devices or in low-bandwidth networks.
    \item Limited generalization performance: Imbalanced datasets with few instances of negative feedback may result in biased reputation scores, as the model may be more likely to assign positive scores to services even if they are not of high quality. 
    \item Difficulty in feature extraction: Imbalanced datasets may make it difficult for the deep learning model to extract meaningful features that accurately represent the characteristics of the service. This can result in poor model performance and inaccurate reputation scores.
\end{itemize}

\section{Proposed Approach}
\begin{figure}
    \centering
    \includegraphics[scale=0.85]{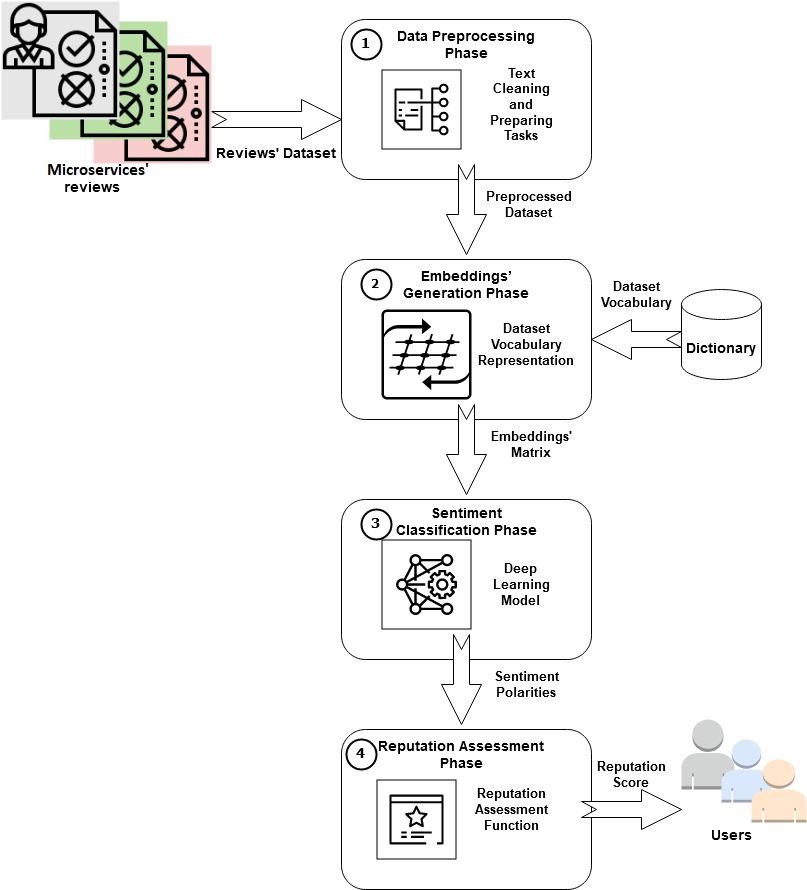}
    \caption{Proposed approach }
    \label{fig:approach}
\end{figure}

Our proposed approach for computing Web services’ reputation focuses on using deep learning models. This choice is justified by the fact that these models have proven their efficiency in sentiment analysis in several applications (i.e., social media monitoring, brand monitoring, market analysis, etc.), as demonstrated in the study presented in \cite{yue2019survey}.  Our approach consists of four phases: 1) the data preprocessing phase, 2) the embedding generation phase, 3) the sentiment analysis phase, and 4) the reputation assessment phase. Figure \ref{fig:approach} presents our approach with its different phases.
\subsection{Data Collecting Phase}
This phase encloses four consecutive tasks, which are:
\begin{enumerate}
    \item Removing the invalid reviews: the reviews’ dataset is examined to filter out invalid reviews. A review is considered invalid if: 1) it is empty, 2) it contains mainly tagged usernames, and 3) it provides mainly commercial URLs.
    \item Word tokenizing and stemming: for each review, tokenization and stemming tasks are performed. Tokenization aims to divide a text into small units called tokens, which refer in our context to words composing the whole review. Stemming aims to reduce a word to its word stem. For example, the stem word of "understanding" is "understand", which is obtained by removing the affix from "understanding".
    \item Stop words, special characters, and punctuation marks removing: stop words such as "a", "of", and "in" are words that need to be filtered out since they do not contribute much to the overall meaning of the review. Also, special characters (i.e., "@", "\%", "/", etc.) and punctuation marks are eliminated to increase the accuracy of the sentiment classification phase.
    \item Part-of-speech (POS) tagging: This task aims to convert each review into a set of tuples where each tuple has a form (word, tag). The tag signifies whether the word is a noun, adjective, verb, etc. After applying POS tagging, only nouns, and adjectives are kept since they both play a key role in the distinction of the sentiment polarity of the review.
\end{enumerate}
\subsection{Embeddings’ Generation Phase}
A word embedding is a learned representation for text where words that have the same meaning have a similar representation. Word embeddings are a class of techniques where individual words are represented as real-valued vectors in a predefined vector space. Each word is mapped to one vector, and the vector values are learned in a way that resembles a neural network. Hence the technique is often lumped into the field of deep learning. To represent the preprocessed data, we proceed with the following successive steps:
\begin{enumerate}
    \item Create a word-to-index dictionary: each word will be assigned to a key, and the unique matching index is used as the value for the key.
    \item Padding: Padding is the process of setting a fixed length to sentences. Every sentence has a different length so we will set the maximum size of each list of sentences to 50 as an example. If the list’s size is greater than 50, it will be trimmed to 50. And for the lists with a length of less than 50, we will add 0 at the end until it reaches the maximum length.
    \item Create a feature matrix: We will load the GloVe word embeddings, which is an algorithm for obtaining vector representations for words. And build a dictionary that will include words as keys and their corresponding embedding list as values.
    \item Create embedding matrix: The matrix will have columns where all columns contain the GloVe word embeddings for the words, and each row will match the corresponding index.
\end{enumerate}
\subsection{Classification Phase}
\begin{figure}
    \centering
    \includegraphics[scale=0.6]{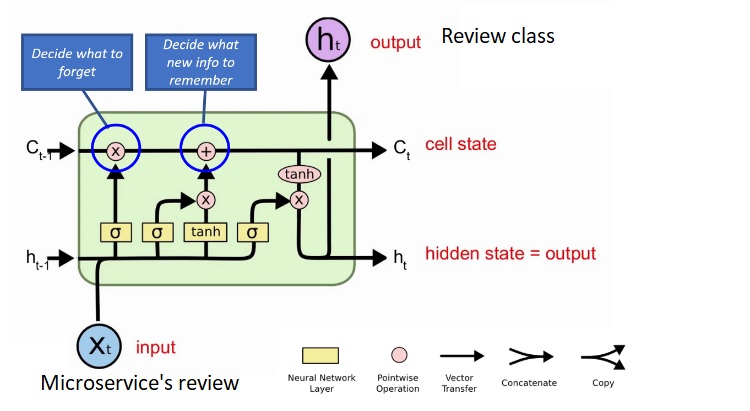}
    \caption{LSTM Architecture for Microservices' Reviews Classification}
    \label{fig:approacharchitecture}
\end{figure}

We propose a deep learning-based sentiment analysis method to ensure review classification. This method relies on the LSTM model. LSTM is a Recurrent Neural Network (RNN) variant specifically designed to better handle long-term dependencies in sequential data. Compared to traditional RNNs, LSTM can selectively forget or remember previous inputs and outputs, allowing it to capture more complex patterns in sequential data. In the context of text classification for sentiment analysis, LSTM can bring several improvements over traditional RNNs:
\begin{itemize}
    \item Better handling of long-term dependencies: Sentiment analysis often requires understanding the context and meaning of words and phrases over long sequences of text. LSTM can better capture these dependencies and make more accurate predictions compared to traditional RNNs.
    \item Improved memory: Since LSTM can selectively remember or forget previous inputs and outputs, it can retain useful information and discard irrelevant information more effectively. This makes it easier for LSTM to identify important features for sentiment analysis and make more accurate predictions.
    \item Reduced vanishing gradient problem: Traditional RNNs can suffer from the vanishing gradient problem, where the gradients become very small, and the model stops learning effectively. LSTM can alleviate this problem by using gating mechanisms to control the flow of information and gradients through the network.
    \end{itemize}
Figure \ref{fig:approacharchitecture} presents the architecture of the LSTM model used for microservices' reviews classification.

\subsection{Reputation Assessment Phase}
The objective of this phase is to use the NBR formula to assess the reputation of Web microservice providers. This will validate the proposed model's effectiveness used for reputation assessment. The NBR formula determines the net value of a brand's reputation based on published reviews, utilizing sentiment analysis to measure customer satisfaction. The NBR index emphasizes positive feedback from brand advocates more than negative feedback, and its output can range from -100 to 100, with higher values indicating a greater number of positive reviews. Equation \ref{eq:1} illustrates the NBR formula.
\begin{equation}\label{eq:1}
NBR = \frac{Positive Reviews - Negative Reviews}{Positive Reviews + Negative Reviews}*100                      
\end{equation}
\section{Experiments}
In this section, firstly, we will present the details of the implementation. Next, we will describe the dataset and the performance metrics. Finally, we will provide a detailed explanation of the results and make comparisons with existing deep-learning models used for text mining.
\subsection{Implementation Environment}
The experiments in this paper are carried out on a PC with the following configuration properties: an x64 CPU, an Intel Core i9-11900H (11th Gen), 32 GB RAM, and an NVIDIA GeForce RTX 3080 (8G) graphics card. All experiments were carried out on Google Colab14, with Python 3.7.1015 and Keras 2.4.3.
\subsection{Dataset}
The reviews are scraped from multiple review websites, including Capterra, g2, Gartner, TrustRadius, Software Advice, GetApp, Trust Pilot,  and Spiceworks. The reviews are about 15 Amazon Web microservices. The collected dataset contains 10,676 reviews, including 10,155 (95\%) “Positive” reviews and 521 (5\%) “Negative” reviews. Duplicates and noises were removed from reviews. Due to the enormous amount of gathered reviews that was processed, manual labeling of this dataset was impracticable. For this reason, we applied a two-stage labeling approach. Firstly, a sentiment analysis technique was utilized to label the dataset automatically. Then, reviews of the minority class were carefully reviewed and re-labeled based on specific features. The dataset was split into 80\% for model training and 20\% for validation and testing. 
\subsection{Performance Metrics}
The overall accuracy performance of the proposed approach is measured through the accuracy, precision, recall, and F1-score, which are expressed in the following:
In order to assess the performance of the proposed approach, accuracy, precision, recall, and F1-score metrics were used.
The statistical measures are represented mathematically in Equations \ref{eq:4} – \ref{eq:8}, where: TP, TN, FP, and FN represent the number of True Positives,
True Negatives, False Positives, and False Negatives, respectively.

\textbf{Accuracy:} it is used to evaluate the model's overall performance throughout all categories. 
\begin{equation}\label{eq:4}
    Accuracy=  \frac{TP+TN}{TP+TN+FP+FN}
\end{equation}
\textbf{Precision:} it is used to assess the model's accuracy in classifying a sample as positive or negative. 
\begin{equation}\label{eq:5}
    Precision=  \frac{TP}{TP+FP}
\end{equation}
\textbf{Recall :} it is employed to assess the model's ability to identify the positive samples. 
\begin{equation}\label{eq:6}
    Recall=  \frac{TP}{TP+FN}
\end{equation}
\textbf{F1-score:} it combines the accuracy and recall measurements to produce a value-added rating for performance verification.
\begin{equation}\label{eq:8}
    F1-score=  \frac{2*Precision*Recall}{Precision+Recall}
\end{equation}

\subsection{Results and Discussion}
The main goal of the proposed approach is to classify microservice reviews properly. This was accomplished using RNN, GRU, CNN, and LSTM. Across 20 epochs, the five deep-learning architectures were trained. The Adam optimizer, the cross-entropy loss function, and the SoftMax activation function have been employed for the models' configuration.

\begin{figure}
    \centering
    \includegraphics[scale=0.6]{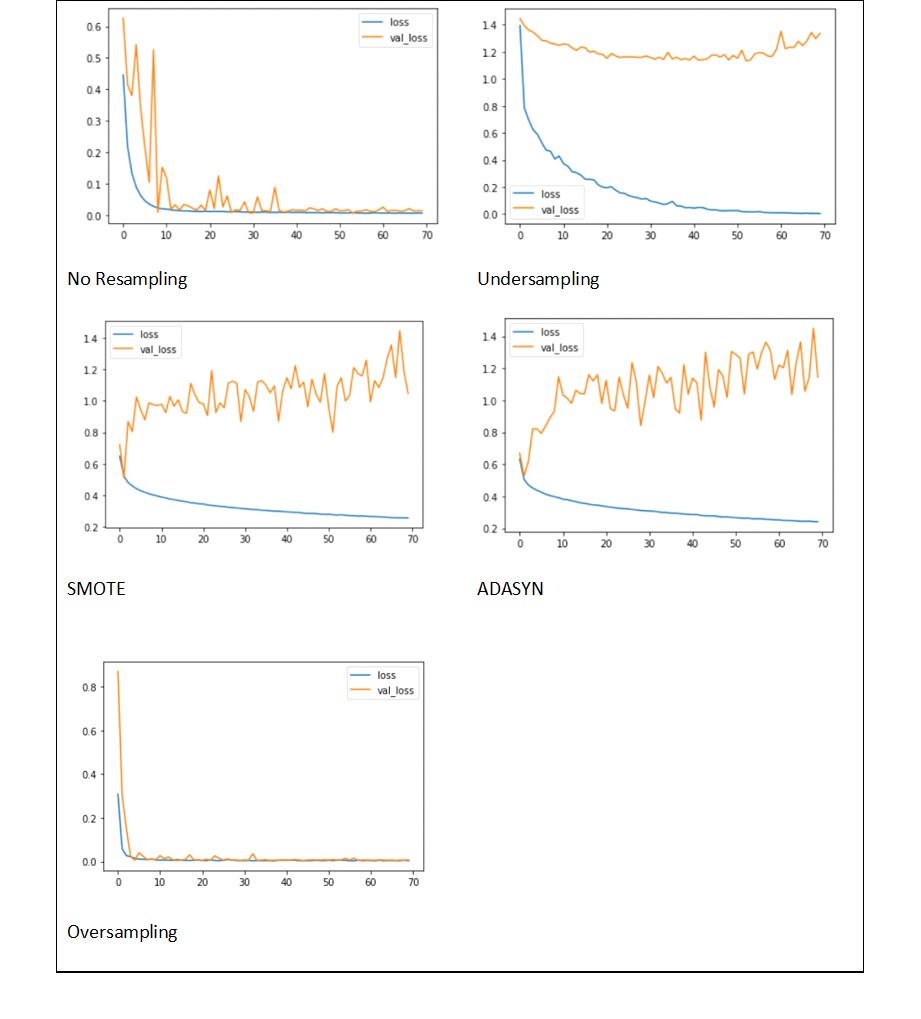}
    \caption{Training Loss and Validation Loss Learning Curves of LSTM Plotted as a Function of the Epoch Number after Applying Different Resampling Strategies.}
    \label{fig:resampling}
\end{figure}

\begin{figure}
    \centering
    \includegraphics[scale=0.65]{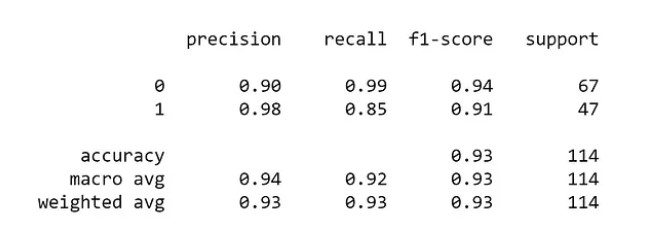}
    \caption{Classification report}
    \label{fig:clf}
\end{figure}

\begin{table}[]
\centering
\caption{Weighted Average Measures of Accuracy, Precision, Recall, and F1-score for RNN, GRU, CNN, and LSTM Models Used for Microservices' Reviews Classification}
\begin{tabular}{|c|l|l|l|c|}
\hline
\textbf{Deep Learning Model} &\textbf{Accuracy (\%)}& \textbf{Precision (\%)} & \textbf{Recall (\%)} & \textbf{F1-score (\%)} \\ \hline
RNN  &    87  &          88          &       86          &     88              \\ \hline
GRU        &  81 &           87       &        90         & 91                  \\ \hline
CNN          &  88 &         92       &            87     &    89              \\ \hline
LSTM (Proposed Model)        & 91 &         92          &  90               &        92           \\ \hline
\end{tabular}
\end{table}
As shown by the performance results in Table 1, our model outperforms all the other models for the weighted average by ensuring an overall accuracy of 91\%, a precision and an F1-score of 92\%, and a recall of 90\%.
The training time for each model is shown in Table 2. The results show that CNN takes the least training time, followed by LSTM. As compared to the training times of RNN and GRU models, the training time of our suggested classifier was acceptable.
\begin{table}[]
\centering
\caption{Training Time for RNN, GRU, CNN, and LSTM Models Used for Microservices' Reviews Classification}
\begin{tabular}{|c|c|}
\hline
\textbf{Deep Learning Model} &\textbf{Training Time (ms)} \\ \hline
RNN  &       698.13         \\ \hline
GRU        &  785.66                  \\ \hline
CNN          &  352.33               \\ \hline
LSTM (Proposed Model)        & 398.41   \\ \hline
\end{tabular}
\end{table}
The considered dataset was a highly imbalanced dataset. It is challenging for any classifier to predict a class accurately based on a few hundred instances. Only 521 negative reviews are included in the whole dataset, with only 104 of them used for testing and validation. To address the imbalance problem, various resampling strategies were tested. These include oversampling, undersampling, SMOTE, and ADASYN strategies. Figure \ref{fig:resampling} shows the training and validation loss learning curves of CBiLSTM with different resampling techniques. All of the other techniques, with the exception of oversampling, appear to be unable to solve the typical underfitting problem during model training. 
Figure \ref{fig:clf} shows the classification report obtained after applying oversampling by considering the same number of positive and negative reviews in the testing, which is 1000. The results confirmed the oversampling strategy's effectiveness since it provided considerable improvements in performance compared to testing results without a resampling strategy. Before oversampling, the model had an accuracy of 91\% and a precision of 92\%. However, after oversampling the data, the model's accuracy and precision increased to 93\%.

The analysis of the binary matrix revealed that the number of positive reviews was 2039, denoted by the TP value, while the number of negative reviews was 112, represented by the TN value. Substituting these values into equation \ref{eq:1}, the NBR score for AWS microservices was computed as 89.58\%. Moreover, the testing dataset comprised 2,031 positive reviews and 104 negative reviews, leading to an estimated reputation score of 90.25\% for AWS microservices. Comparing the NBR score generated using LSTM-based techniques for reputation assessment with the score obtained from the original dataset revealed close similarity between the two values. These results imply that the LSTM-based approach can be a reliable and effective technique for assessing the reputation of microservices providers.
%
%
\section{Conclusion and Future Work}
This study develops a deep learning model to classify web microservice user-related reviews based on sentiments derived from collected users' reviews. The proposed deep learning model, LSTM, outperforms other existing models used in text classification for sentiment analysis, such as RNN, GRU, and CNN. The aim of this approach is to establish a reputation ranking for microservices providers by analyzing the QoE of their users. The QoE is gauged by classifying reviews as "positive" or "negative" and comprehensively evaluating users' opinions towards the service providers. Our upcoming work involves the integration of advanced natural language processing techniques to enhance the precision of sentiment analysis. This may entail the use of sophisticated deep learning models, like Transformers, that leverage attention mechanisms to more effectively comprehend the nuances of language and context in reviews. Additionally, we will investigate the impact of the suspicious user punishment mechanism on the reputation of service providers and we will propose viable solutions to address the challenges posed by unjust feedback ratings.

\section*{Acknowledgment}
The author would like to thank Prince Sultan University for financially supporting the conference registration fees.
%
%
\bibliographystyle{splncs04}  
\bibliography{refs}  

\begin{thebibliography}{10}
\providecommand{\url}[1]{\texttt{#1}}
\providecommand{\urlprefix}{URL }
\providecommand{\doi}[1]{https://doi.org/#1}

\bibitem{al2022cbilstm}
Al~Saleh, R., Driss, M., Almomani, I.: Cbilstm: A hybrid deep learning model
  for efficient reputation assessment of cloud services. IEEE Access
  \textbf{10},  35321--35335 (2022)

\bibitem{baldick1996concise}
Baldick, C.: The concise Oxford dictionary of literary terms. Oxford University
  Press (1996)

\bibitem{Bilecki2017trust}
Bilecki, L.F., Fiorese, A.: A trust reputation architecture for cloud computing
  environment. In: 2017 IEEE/ACS 14th International Conference on Computer
  Systems and Applications (AICCSA). pp. 614--621. IEEE (2017)

\bibitem{BIRJALI2021Comprehensive}
Birjali, M., Kasri, M., Beni-Hssane, A.: A comprehensive survey on sentiment
  analysis: Approaches, challenges and trends. Knowledge-Based Systems
  \textbf{226},  107134 (2021)

\bibitem{driss2022ws}
Driss, M.: Ws-advising: a reusable and reconfigurable microservices-based
  platform for effective academic advising. Journal of Ambient Intelligence and
  Humanized Computing pp. 1--12 (2022)

\bibitem{driss2022req}
Driss, M., Ben~Atitallah, S., Albalawi, A., Boulila, W.: Req-wscomposer: a
  novel platform for requirements-driven composition of semantic web services.
  Journal of Ambient Intelligence and Humanized Computing pp. 1--17 (2022)

\bibitem{hasnain2022machine}
Hasnain, M., Ghani, I., Pasha, M.F., Jeong, S.R.: Machine learning methods for
  trust-based selection of web services. KSII Transactions on Internet and
  Information Systems (TIIS)  \textbf{16}(1),  38--59 (2022)

\bibitem{hasnain2020evaluating}
Hasnain, M., Pasha, M.F., Ghani, I., Imran, M., Alzahrani, M.Y., Budiarto, R.:
  Evaluating trust prediction and confusion matrix measures for web services
  ranking. IEEE Access  \textbf{8},  90847--90861 (2020)

\bibitem{papadakis2019collaborative}
Papadakis-Vlachopapadopoulos, K., Gonz{\'a}lez, R.S., Dimolitsas, I.,
  Dechouniotis, D., Ferrer, A.J., Papavassiliou, S.: Collaborative sla and
  reputation-based trust management in cloud federations. Future Generation
  Computer Systems  \textbf{100},  498--512 (2019)

\bibitem{saberi2017sentiment}
Saberi, B., Saad, S.: Sentiment analysis or opinion mining: A review. Int. J.
  Adv. Sci. Eng. Inf. Technol  \textbf{7}(5),  1660--1666 (2017)

\bibitem{surianarayanan2019essentials}
Surianarayanan, C., Ganapathy, G., Pethuru, R.: Essentials of Microservices
  Architecture: Paradigms, Applications, and Techniques. Taylor \& Francis
  (2019)

\bibitem{wahab2015survey}
Wahab, O.A., Bentahar, J., Otrok, H., Mourad, A.: A survey on trust and
  reputation models for web services: Single, composite, and communities.
  Decision Support Systems  \textbf{74},  121--134 (2015)

\bibitem{wolff2016microservices}
Wolff, E.: Microservices: flexible software architecture. Addison-Wesley
  Professional (2016)

\bibitem{Yu2019review}
Yu, Y., Si, X., Hu, C., Zhang, J.: A review of recurrent neural networks: Lstm
  cells and network architectures. Neural computation  \textbf{31}(7),
  1235--1270 (2019)

\bibitem{yue2019survey}
Yue, L., Chen, W., Li, X., Zuo, W., Yin, M.: A survey of sentiment analysis in
  social media. Knowledge and Information Systems  \textbf{60},  617--663
  (2019)

\end{thebibliography}
\end{document}